\documentclass[printer]{aa}
\usepackage{graphicx} 
\usepackage{txfonts}
\usepackage{multirow}
\usepackage{lscape}
\usepackage{longtable}
\begin{document} 

\title{EDIBLES II. On the detectability of C$_{60}^{+}$ bands}

\titlerunning{EDIBLES II. C$_{60}^{+}$ diffuse interstellar bands}

\author{R. Lallement\inst{1}
        \and
        N. L.J. Cox\inst{2} 
       \and
        J. Cami\inst{3} 
         \and
        J. Smoker\inst{4} 
        \and
        A. Fahrang\inst{3,5}
        \and
        M. Elyajouri\inst{1}
       \and
       M. A. Cordiner \inst{6,7}
       \and
       H. Linnartz \inst{8}
       \and
       K.T. Smith \inst{9}
       \and
       P. Ehrenfreund \inst{10}
       \and
       B.H. Foing \inst{11}
       }

\authorrunning{Lallement et al}

\institute{GEPI, Observatoire de Paris, PSL Research University, CNRS,
Place Jules Janssen, 92190 Meudon, France\\ \email{rosine.lallement@obspm.fr}
\and
ACRI-ST, 260 route du Pin Montard, 06904, Sophia Antipolis, France
\and
Department of Physics and Astronomy, The University of Western
Ontario, London, ON N6A 3K7, Canada
\and
European Southern Observatory, Alonso de Cordova 3107, Vitacura,
Santiago, Chile
\and
School of Astronomy, Institute for Research in Fundamental Sciences,
19395-5531 Tehran, Iran
\and
NASA Goddard Space Flight Center, 8800 Greenbelt Road, Greenbelt, MD 20771, USA
\and
Department of Physics, Catholic University of America, Washington, DC 20064, USA
\and
Sackler Laboratory for Astrophysics, Leiden Observatory, Leiden
University, PO Box 9513, NL2300 RA Leiden, The Netherlands
\and
AAAS Science International, Clarendon House, Clarendon Road,
Cambridge CB2 8FH, UK
\and
George Washington University, Washington DC, US
\and
ESTEC, ESA, Noordwijk, The Netherlands
}

\date{Received Jan 15, 2018; accepted Jan 29, 2018}

\abstract{Gas phase spectroscopic laboratory experiments for the buckminsterfullerene cation C$_{60}^{+}$ resulted in accurate rest wavelengths for five C$_{60}^{+}$ transitions  that have been compared with diffuse interstellar bands (DIBs) in the near infra-red. Detecting these in astronomical spectra is difficult due to the strong contamination of ground-based spectra by atmospheric water vapor, to the presence of weak and shallow stellar lines and/or blending with other weak DIBs. The detection of the two strong bands 
has been claimed by several teams, and the three additional and weaker bands have been detected in a few sources. Certain recent papers have argued against the  identification of C$_{60}^{+}$ based on spectral analyses claiming (i) a large variation in the ratio between the equivalent widths of the 9632 and 9577\AA\: bands, (ii) a large redshift of the 9632\AA\: band for the Orion star HD\,37022, and (iii) the non-detection of the weaker 9428\AA~DIB. Here we address these three points. (i) We show that the model stellar line correction for the 9632\AA~DIB overestimates the difference between the strengths of the lines in giant and dwarf star spectra, casting doubts on the conclusions about the ratio variability. (ii) Using high quality stellar spectra from the ESO Diffuse Interstellar Bands Large Exploration Survey (EDIBLES), recorded with the ESO/Paranal Ultraviolet Echelle Spectrograph (UVES)  in about the same atmospheric conditions, we find no wavelength shift in the 9632\AA\ band towards HD\,37022. (iii) Using EDIBLES spectra and data from the Echelle SpectroPolarimetric Device for the Observation of Stars (ESPaDOnS) at CFHT we show that the presence of a weak 9428\AA\ band cannot be ruled out, even in the same observations that a previous study claimed it was not present.
}


\maketitle

\section{Introduction}

One of the longest standing spectroscopic mysteries in interstellar medium (ISM) studies is the identity of the carriers of the so-called diffuse interstellar bands (DIBs) \citep[see e.g.][and references therein]{herbig95,Sarre06,Snow14,CamiCox14}. Proposed candidate DIB carriers are large carbonaceous molecules in the gaseous phase, such as polycyclic aromatic hydrocarbons (PAHs) and their cations, polyhedral carbon ions, or very complex hetero-cyclic aromatic-rich moieties  \citep[see e.g.][]{Leger85,Vanderzwet85, Leger88, Jones_A_16,Omont16}. None of the hundreds of optical or infra-red (IR) bands has been fully and unambiguously attributed to a given species. A remarkable case, however, is the one of near-infrared (NIR) bands due to the C$_{60}^{+}$ cation. C$_{60}^{+}$ and its parent molecule C$_{60}$ have been definitely detected in emission in circumstellar and interstellar environments \citep{Cami10, Sellgren10, Berne13}. C$_{60}^{+}$ was detected for the first time in absorption by \cite{Foing94} based on similarities with laboratory neon matrix spectra \citep{Fulara93}. The criteria for C$_{60}^{+}$ identification were discussed in detail in \cite{Ehrenfreund95}. According to \cite{Bendale92}, the ground state of C$_{60}^{+}$ departs from I$_{h}$ symmetry of neutral C$_{60}$ towards the D$_{5d}$  geometry,  with a Jahn-Teller distortion stabilization
energy of 8.1 kcal mol$^{-1}$. The confirmations of the two strong bands have been reported by many authors \citep{Foing97,Jenniskens97,Gala00, Cox14, Hamano15, Walker15, Hamano16, Walker16, Walker17} and detections of between one and three of the weaker absorption bands have been made towards a few sources \citep{Walker15,Walker16}, but discrepancies between predicted and observed wavelengths and intensity ratios have been also reported \citep{Jenniskens97, Gala17, GalaKre17}. 

On the experimental side, new laboratory spectra of C$_{60}^{+}$-He complexes and C$_{60}^{+}$ embedded in He-droplets have confirmed  the existence of five bands in the NIR and brought additional precision in their characterization \citep{Campbell15, Campbell16b, Campbell16a, Kuhn16, Spieler17}. The detection of the NIR C$_{60}^{+}$ bands from ground-based data is difficult. Firstly, the spectral regions of the five DIBs are very strongly contaminated by telluric water vapor lines. Secondly, weak stellar lines are present in the spectra of the early-type stars used for the DIB detections. In the infrared, their widths and depths can be very similar to those of the DIBs and they may overlap and mimic the weak DIBs \citep{Foing94,Cordiner17,Gala17}. Finally, additional weak DIBs are also partially blended with the expected bands \citep{Walker15, Walker16,Walker17,GalaKre17}. Very recently, a new technique using the Space Telescope Imaging Spectrograph (STIS) on board the Hubble Space Telescope (HST) has been developed by \cite{Cordiner17} and brought promising results. Such data that are free of telluric contamination, are expected to close the controversy about the C$_{60}^{+}$ identification.

While some of the previously mentioned observations of reddened stars and their analyses found a convincing match between laboratory data and observed bands within observational uncertainties \citep{Walker15, Walker16,Walker17}, other recent studies based on the same high quality, high resolution spectra have questioned some of the claimed detections \citep{Gala17,GalaKre17}. In Sect.~\ref{sec:2}, we revisit the 9632/9577 DIB ratio study and the correction for contaminating stellar lines that are part of \cite{Gala17}. In Sect.~\ref{sec:3}, we revisit the double structure of the 9632~\AA\ DIB of the Orion star \object{HD\,37022}, making use of new UVES spectra of the star and two additional targets. In Sect.~\ref{sec:4}, we revisit the suggested absence of the weak DIB around 9428~\AA.  Firstly we use the ESPaDOnS spectra of \object{Cyg OB2 5} (\object{BD +40 1220}), \object{Cyg OB2 12} and a weakly reddened target. Secondly, we use a new UVES spectrum of \object{HD\,169454} as well as the UVES spectra of the same two additional targets from Sect.~\ref{sec:3}. We conclude and discuss these analyses in Sect.~\ref{sec:5}.

\begin{figure}[t!]
  \centering
   \includegraphics[width=0.99\hsize]{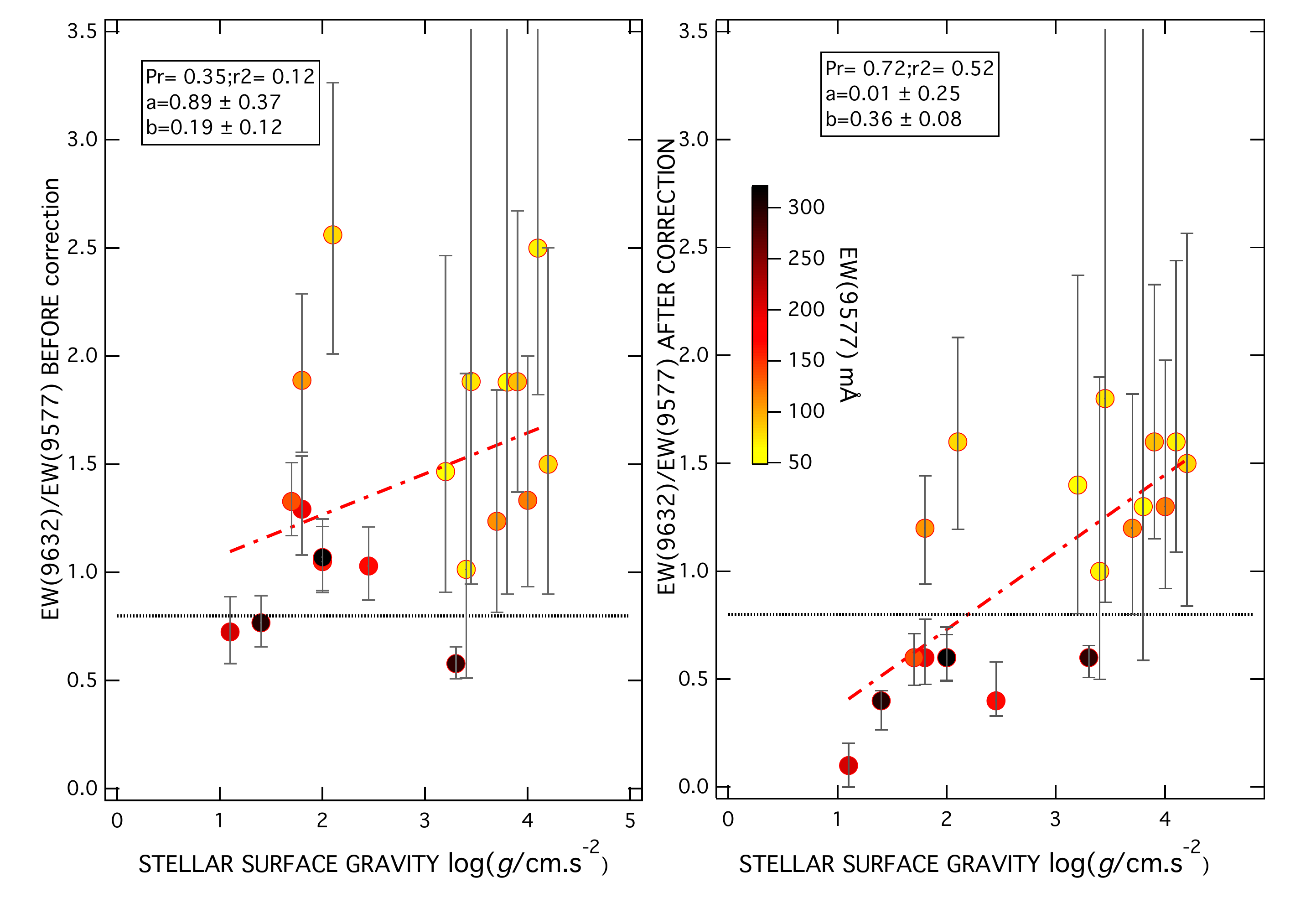}
      \caption{Ratio between the equivalent widths (EWs) of the 9632 and 9577~\AA\  absorptions, before (left) and after (right) removal of the stellar Mg\,{\sc ii} line that contaminates the latter. EWs are from table~2 of \cite{Gala17} and the ratio is displayed as a function of the target star surface gravity represented by the log($g$) parameter. Parameters a (intercept) and b (gradient) of a linear fit are listed in the graph as well as the Pearson (Pr) and r-squared (r$_{2}$) correlation coefficients. The horizontal line indicates the EW ratio derived from the laboratory data.}
         \label{dibratio}
   \end{figure}

\section{9632~\AA\ stellar contamination and the 9632/9577 DIB ratio}\label{sec:2}

Based on high resolution, high signal observations of reddened stars recorded in excellent atmospheric conditions, \cite{Gala17} and \cite{GalaKre17} have studied in detail the spectral regions around the expected C$_{60}^{+}$  9577 and 9632~\AA\ transitions. In particular,  \cite{Gala17} have modeled the stellar Mg\,{\sc ii} (Mg$^{+}$) line that contaminates the 9632~\AA\ DIB and have computed the ratios between the two DIB equivalent widths along 19 line-of-sights, both before and after removal of the contaminating stellar line. They found that the corrected 9632/9577 DIB ratios range from a factor of at least 6 from 0.2 to 1.2, largely different from the $\sim$0.8 value derived from laboratory work and with a significantly larger variability than what one would expect for bands originating from one and the same carrier. The authors therefore concluded that C$_{60}^{+}$ cannot be the carrier of these two bands. In laboratory experiments, ratios of bands in the spectra of tagged molecules vary under experimental conditions \citep{Fulara93,Pino99}, and in the case of C$_{60}^{+}$ this variability has been pointed out soon after the first DIB detection \citep{Ehrenfreund95,Ehrenfreund97}. The 9632/9577 ratio has been found to vary in response to the nature of the atom or molecule attached to C$_{60}^{+}$ and to temperature variations \citep{Campbell15,Campbell16b,Campbell16a,Holz17,Spieler17}, therefore astronomical measurements are not expected to match perfectly any of them \citep[see, e.g. ][]{Walker17}. As measured by \cite{Holz17}, the use of Ne, Ar, Kr, H$_{2}$, D$_{2}$ or N$_{2}$ as ligand with C$_{60}^{+}$ has the effect of splitting the 10,378 cm$^{-1}$ line (corresponding to the 9632~\AA\ absorption), and the line ratio variability is due to changes in the population of the $^{2}$E$_{1(2)u}$ excited electronic state(s) lying a few cm$^{-1}$ above the $^{2}$H$_{u}$ $^{2}$A$_{1u}$ level (in the D${5d}$ geometry).  According to \cite{Holz17}, He produces the smallest perturbations, and indeed there is no observed splitting when the ligand is He. On the other hand, laser dissociation spectroscopy of C$_{60}^{+}$ embedded in ultra-cold He droplets suggests that the intensity ratio for the bare C$_{60}^{+}$ bands will not differ much from that derived experimentally for He$_{n}$-C$_{60}^{+}$ and the lower values of the number n of attached He atoms \citep{Spieler17}. For those reasons, we assume in what follows that the most probable 9632/9577 ratio in the ISM must be close to the one deduced from fragmentation spectra of C$_{60}^{+}$-He in cryogenic ion-trap experiments \citep{Campbell16b,Campbell16a}, i.e. 0.8 \citep[with an estimated uncertainty of 20\%, see ][]{Walker17} and maybe constant. We note, however, that this remains an assumption, since none of the experiments has been done at temperatures above $\sim$ 10K, while temperatures measured in diffuse clouds may be higher.

We have performed several tests of a potential residual influence of the stellar atmospheric parameters on the correction performed by \cite{Gala17}. To do so we have used the equivalent widths (EWs) of the two DIBs listed in their table 2. Fig.~\ref{dibratio} displays the EW ratios before and after the correction, as a function of the stellar surface gravity, represented by the log($g$) parameter (listed in \cite{Gala17} table 1). Error bars shown in the Figure are computed based on the minimum and maximum values of the equivalent width ratios, using errors on each EW as quoted in the tables. The colour scale refers to the 9577\AA\ DIB strength. As can be seen in the Figure, before the correction the 9632/9577 EW ratio is strongly scattered and it is predominantly above the expected laboratory ratio of 0.8, which leaves room for a decrease down to this ratio after removal for a stellar contribution to the equivalent width of the 9632~\AA\ DIB. We also note a small positive dependence on the star surface gravity, a trend that is not unexpected if there is a stellar contamination, but may also be due to the fact that the majority of the weaker DIBs corresponds to observations of dwarf targets (see the colour scale). After the correction, an ascending trend is clearly visible, showing that the new ratio using the corrected values presented by \cite{Gala17} is significantly smaller in the case of giant stars than in the case of dwarfs. This is confirmed by coefficients and corresponding errors of a  linear fit to the EW ratio as a function of log($g$), whose results are displayed in the Figure, and by the corresponding Pearson and r-squared correlation coefficients. After the correction the positive gradient 0.36 $\pm$ 0.08 is about four times its uncertainty deduced from the mean data point dispersion. 

There are two ways to interpret this trend. One is that either the 9577~\AA\ or the corrected 9632~\AA\ absorption is stellar in nature and strongly depends on the stellar type. However, both absorptions only appear in highly reddening star spectra, and this first hypothesis is therefore very unlikely. A second one is that the equivalent width of the contaminating stellar line has been overestimated for giant stars and (or) underestimated for dwarfs. In this case a correction for such a bias leaves room for much less variability or constancy of the ratio, and one can see from the figure that the predicted value of 0.8 can not be excluded. This is in agreement with the conclusions of \cite{Walker16} and \cite{Walker17} based on the use of close spectral standards to correct for the stellar contamination. In any case, this shows that deeper studies of the stellar contamination are still necessary to determine with high accuracy the 9632/9577 ratio and preclude or confirm its constancy.

\section{The 9632\AA\ DIB towards HD\,37022}\label{sec:3}

\begin{figure}[t!]
  \centering
   \includegraphics[width=0.99\hsize]{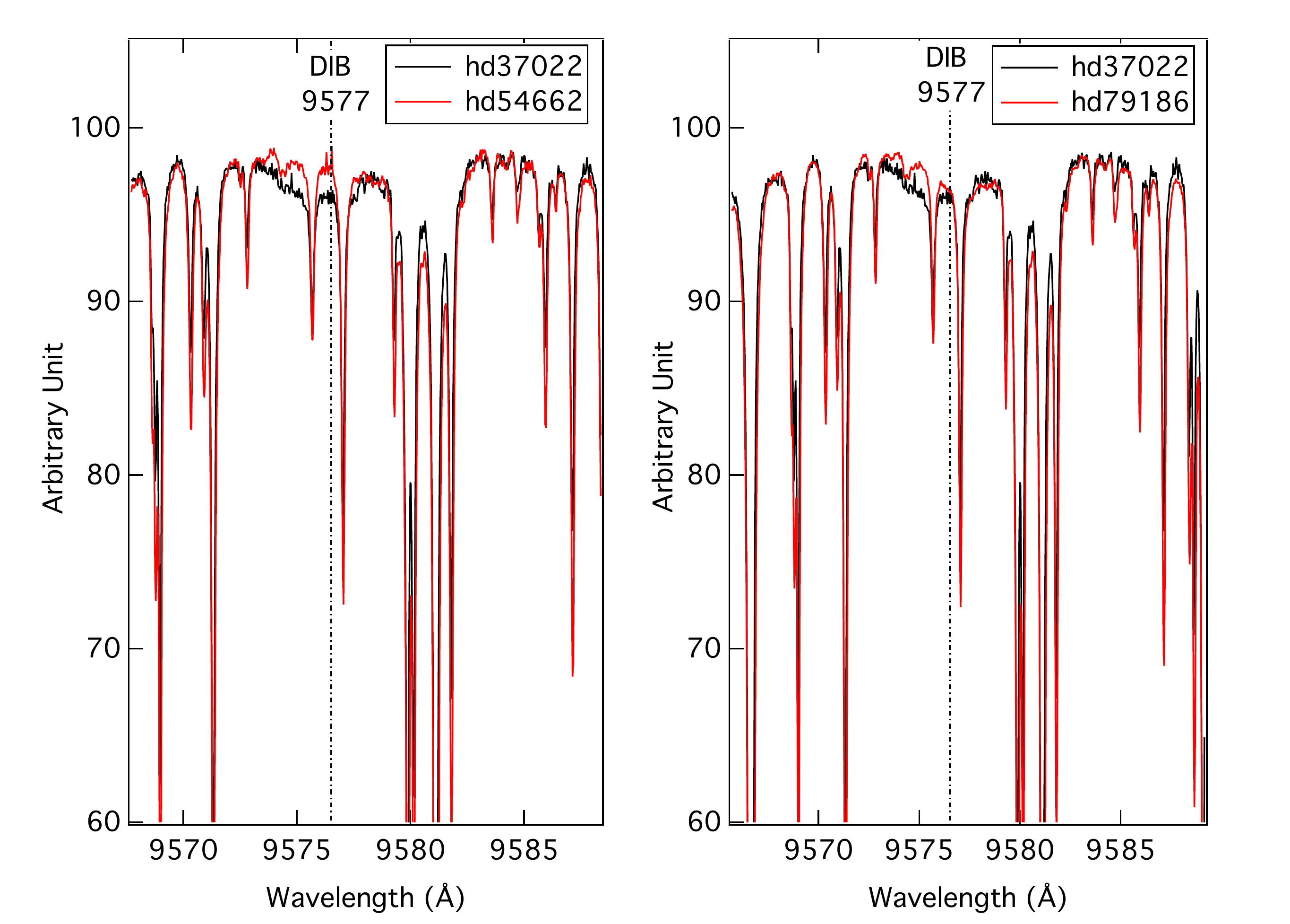}
      \caption{Comparisons between the 9577~\AA\ DIB in HD\,37022 and the one in HD\,54662 (left) and HD\,79186 (right). The depth of the HD\,37022 DIB is much larger than for the two other stars which have a similar reddening.}
         \label{stars37022_54662_9577}
   \end{figure}

\begin{figure}[t!]
  \centering
 \includegraphics[width=0.99\hsize]{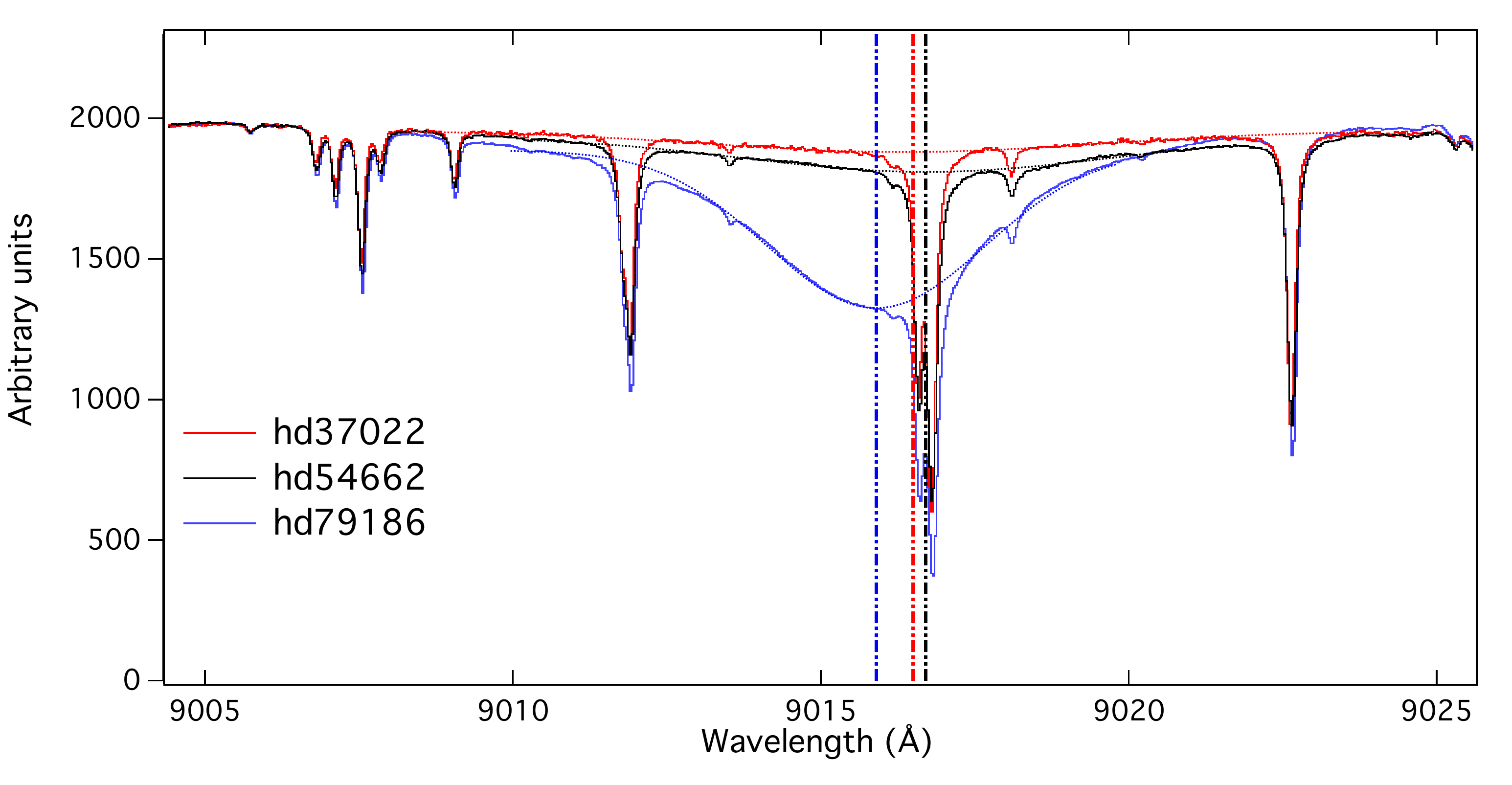}
   \caption{One of the stellar lines present in the same spectral region as the 9632~\AA\ DIB. The spectra are in the geocentric frame so the telluric lines coincide. The three stellar lines are fitted with Gaussian profiles and the ordering of their centers will be used in Fig.~\ref{stars3_9632} (see text).}
         \label{stars3stell}
   \end{figure}

\begin{figure}[t!]
  \centering
 \includegraphics[width=0.99\hsize]{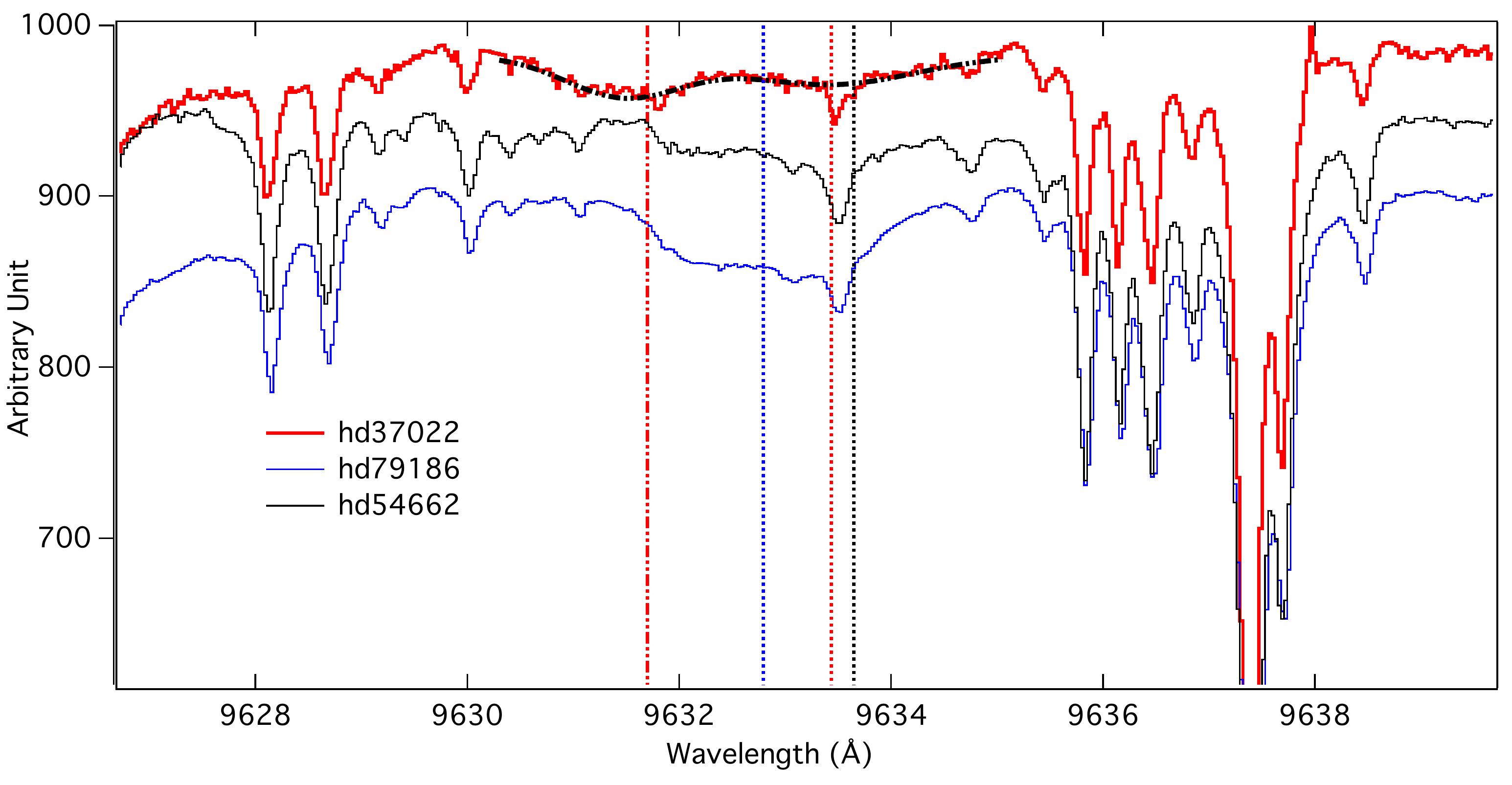}
  \caption{The same three recorded spectra as in Fig.~\ref{stars3stell}, in the geocentric frame, in the region of the 9632~\AA\ DIB. A vertical offset has been applied to each spectrum for clarity. The HD\,37022 spectrum exhibits a clear double absorption, as  first observed by \cite{Gala17}, here delineated by two Gaussian profiles (thick black dashed line). The expected center of the 9632~\AA\ DIB, based on the 9577~\AA\ absorption, or the Na\,{\sc ii} or K\,{\sc i}  lines, is shown by a red dot-dot-dashed line and corresponds very well to the center of the  blue-ward absorption. The center of the second absorption is shown by a red dotted line. If this absorption is stellar, the corresponding stellar lines of the two other targets should be centered at the black (HD\,54662) and blue (HD\,79186) dotted line locations, using the result of Fig.~\ref{stars3stell}. In both cases these fall in the middle of a broad absorption, suggesting that all three absorptions maybe stellar.}
         \label{stars3_9632}
   \end{figure}

The two strong C$_{60}^{+}$ bands have been observed towards the Orion Trapezium HD 37022 \citep{Ehrenfreund97}. The two DIBs increase in this region dominated by UV radiation, as expected for the C$_{60}$ molecule ionization/recombination properties. The low ionization potential of C$_{60}$ (7.61 eV) favours its ionization in the diffuse medium outside dark clouds, a trend demonstrated  by \cite{Cordiner17}.
\cite{Krelowski15} and \cite{Gala17} have analyzed the 9577 and 9632~\AA\ DIBs in more recent spectra of this star. \cite{Krelowski15} found that some of the optical DIBs are shifted w.r.t. others by up to 10-20~km~s$^{-1}$ and that some DIB ratios are very different from those measured in general.  In the case of the C$_{60}^{+}$ bands, \cite{Gala17} have shown that, in the 9632~\AA\ DIB region, there are two clearly spectrally separated absorptions (see their Fig.~8). They attributed the short wavelength absorption to the contaminating stellar Mg\,{\sc ii} line discussed in the previous section, and the second one to the interstellar counterpart. Doing so, and referring to experimental wavelengths, they derived a strong discrepancy between the radial velocities of the two DIBs, the 9632~\AA\ DIB being redshifted by 48~km~s$^{-1}$ with respect to the 9577~\AA\ absorption or equivalently w.r.t. the expected position derived on the interstellar neutral potassium line. This finding was taken as a further argument against C$_{60}^{+}$ as a carrier of these NIR DIBs.

Shifts of DIBs due to different species may exist if the sightline contains several clouds at well separated radial velocities and the clouds have very different abundances of DIB carriers. \cite{Cami97} have shown evidence that DIBs behave differently in Orion, compared to other less-UV-irradiated environments, and the ionization state of the various clouds observed towards the Orion stars may explain at least partially the existence of Doppler shifts between the DIBs. As a matter of fact, the clouds observed towards HD\,37022 and traced by the main species \ion{Na}{i}, \ion{Ca}{ii} and \ion{K}{i}, have radial velocities ranging from +5 to +40~km~s$^{-1}$ \citep[see ][]{Krelowski15}. The cold neutral clouds traced by \ion{Na}{i} and \ion{K}{i} have the lowest radial velocities, while the most ionized clouds traced by \ion{Ca}{ii} are redshifted. In a recent principal component analysis, \cite{Ensor17} have demonstrated the \textit{hierarchical} behavior of optical DIBs w.r.t. the radiation field, extending to a large set of strong DIBs the well-known \textit{skin effect} observed for the 5780-5797~\AA\ pair. This is also in agreement with a recent analysis by \cite{ElYajouri17} for five of the DIBs in the 5770-6620 ~\AA\ region. Interestingly, the redshifts observed by \cite{Krelowski15} are found for the two strong DIBs that are the most favoured in high irradiation clouds, namely the 5780 and 6284~\AA\ DIBs, while the less influenced 6196~\AA\ DIB is slightly shifted. We suggest that the DIB carrier columns of the \textit{high irradiation} DIBs are much stronger in the redshifted clouds and are responsible for some of the observed departures.

Such effects, however, can not explain a shift among DIBs from the same species and the result on the 9577, 9632~\AA\ C$_{60}^{+}$ bands. In this work, we have revisited the case of HD\,37022 using recent VLT/UVES spectra of HD\,37022 acquired as part of the EDIBLES survey \citep{CoxEDIBLES}. EDIBLES spectra of two additional targets, \object{HD\,54662} and \object{HD\,79186}, are also used for comparison. Details on the observations, data reduction and targets can be found in \cite{CoxEDIBLES}. We have selected three exposures based on the weakness and rare similarity of the telluric water vapor lines. The precipitable water vapour (PWV) at the time of the HD\,54662 and HD\,79186 observations was 1.34 and 1.51 mm respectively, according to the ESO Paranal radiometer, well below average conditions. No data was available for the HD\,37022 observation. However, we also checked by eye all EDIBLES  exposures and could easily select the low PWV conditions based on the depths of the lines. 
We have used the spectra of the three targets to investigate further the locations of the stellar and interstellar absorptions around the two  strong C$_{60}^{+}$ bands. HD\,54662 and HD\,79186 are characterized by E(B-V)= 0.32 and 0.28 respectively, similar to the colour excess of 0.31 for HD\,37022 but their 9577~\AA\ bands are significantly weaker than for the peculiar HD\,37022 that is known for its particularly strong absorptions. This difference and the quality of the data are illustrated in Fig.~\ref{stars37022_54662_9577}.

We have used the UVES data to measure the radial velocity differences between the stellar absorptions for the three stars. One example is shown in Fig.~\ref{stars3stell}. In order to facilitate the comparisons we have used the observed spectra (i.e. in the geocentric frame), because the telluric lines are concentrated in the same areas. We have used masks of the telluric lines and Gaussian fits to derive the stellar line centers (as shown in the Figure). These values have allowed us to derive the expected corresponding differences in wavelength, this time for the \ion{Mg}{ii} 9632~\AA\ contaminating lines. In parallel, we have measured the equivalent widths of the 4481~\AA\ \ion{Mg}{ii} line for the three stars, found to be $\sim$ 410, 43 and 35~m\AA\ for HD\,79186, HD\,54662 and HD\,37022 respectively. In Fig.~\ref{stars3_9632} we show the observed spectra, again in the geocentric frame, in the 9632~\AA\ spectral region. The HD\,54662 spectrum has been multiplied by a linear function to correct for the local intensity gradient, in such a way all three spectra have flat continua over the shown spectral range. We clearly see that, as observed by \cite{Gala17}, in the case of HD\,37022 there are two broad absorptions, here fitted by two Gaussian profiles. This is at variance with the two other targets for which only one broad absorption is seen. Assuming, like \cite{Gala17}, that the 9632\AA\ DIB should be centered at the measured interstellar K\,{\sc i} radial velocity, we have estimated the expected location of the 9632~\AA\ DIB using the most recent laboratory wavelength of \cite{Spieler17} and the geocentric-heliocentric correction for the HD\,37022 observation. This expected location is shown in Fig.~\ref{stars3_9632} and falls in the central part of the first (blue-ward) absorption. If this is actually the C$_{60}^{+}$ band, then this derived wavelength is fully consistent with the prediction from the 9577~\AA\  band, found by \cite{Gala17} to be at the average \ion{Na}{i} velocity (as confirmed  by the UVES data). 

The center of the second (red-ward) absorption toward HD\,37022 has been also analyzed using a Gaussian fit and is also included in Fig.~\ref{stars3_9632}. If we assume that its origin is stellar, then we can use its location and the wavelength differences discussed above (and taken from Fig.~\ref{stars3stell}) to locate the corresponding stellar lines for the two other targets. What we found is that these locations agree with the observed centers of the broad absorptions measured towards the two other targets. Since their interstellar absorptions are expected to be significantly weaker than for HD\,37022, these redshifted absorptions are very likely stellar. Accordingly, there are three features distributed in the same way as in the case of the other stellar lines, with two being very likely stellar. A natural interpretation is that the HD\,37022 red-ward absorption is also stellar, while the first is interstellar, in agreement with the 9577 absorption and the other species. This interpretation is reinforced by the observed relative strengths of the three red-ward absorptions, in good agreement with the relative strengths of the 4481~\AA\ \ion{Mg}{ii} lines, namely a stronger line for HD\,79186 and much weaker and similar lines for HD\,37022 and HD\,54662. This is at variance with the interpretation of \cite{Gala17} who have assumed that the blue-ward absorption is stellar and the red-ward interstellar, based on the redshifts of the optical DIBs. Again here, further stellar studies are needed, but the conclusion is that there exists no fully demonstrated discrepancy in Doppler shift for the two stronger C$_{60}^{+}$ bands observed towards HD37022.

\section{Search for the 9428~\AA\ DIB}\label{sec:4}

As already mentioned, telluric lines render the identification of the weak DIBs extremely difficult. An illustration of this difficulty is the discrepancy among results obtained for the same targets, and even for the same exposures. As a recent example, \cite{Walker16} and \cite{Gala17} obtained different results for the same CFHT/ESPaDOnS spectrum of the target star HD\,169454, which resulted in one group claiming a DIB feature at 9428~\AA\ , whereas the other group presented its absence. As recently outlined by \cite{Walker17}, a major problem is linked to the definition of the actual stellar continuum. Even in the absence of saturation, when the atmospheric lines are deep and numerous, the true continuum is everywhere above the spectrum and one can not use telluric-free regions for its measurement. Moreover, even small wings in the instrumental function have a strong effect and increase the gap between the true continuum and the maxima in the spectrum. This can be seen e.g. in Fig.~5 from \cite{Bertaux14_tapas}, or in Fig.~2c from \cite{Cordiner17}. In this latter figure the data has been offset down by 0.1 for clarity, but it can be seen by eye that removing the offset would keep the data below the retrieved continuum. Unless atmospheric temperature and pressure profiles as well as instrumental functions are identical, lacking knowledge of the true continuum prevents the use of the division by the spectrum of a comparison star, because any way to re-scale the depths of the telluric features to adapt to this standard target implicitly hypothesizes a continuum. In addition to the continuum definition, the spectra of early-type giant stars such as those used as  reddened targets may contain  features associated with stellar winds. Such features are easily seen in the case of strong stellar lines, but may be present with very small amplitude at the location of weaker lines. If such broad features are adjacent or blended with the DIBs they may influence the continuum fitting and the normalization. This is why unnormalized spectra must be used and closely inspected before any conclusion is drawn. 
\begin{figure}[t!]
  \centering
   \includegraphics[width=0.99\hsize]{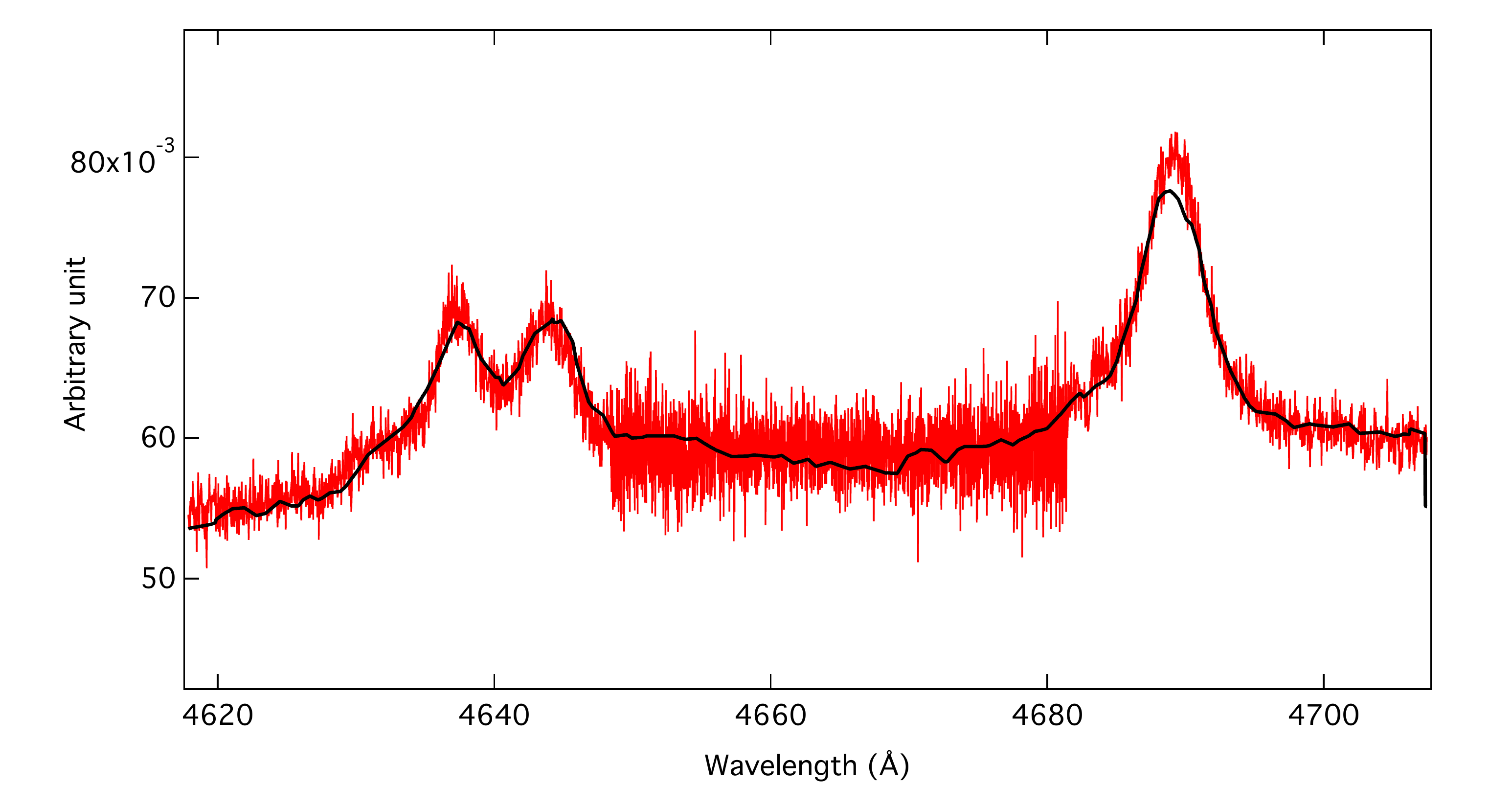}
      \caption{ESPaDOnS spectrum of BD+40 4220 (red line) in the region of the three 4634 and 4640~\AA\  \ion{N}{iii} and 4686~\AA\ \ion{He}{ii} stellar wind emission lines. Superimposed is an average of the two spectra that have been previously observed by \cite{Rauw99} for orbital phases 0.178 and 0.193 of the binary system. A Doppler shift of $+$216~km~s$^{-1}$ has to be applied to match the ESPaDOnS data, a value in agreement with the average orbital phase of 0.185. The same Doppler shift will be applied to the two 9402.5 and  9424.5~\AA\ \ion{N}{iii} stellar wind emission lines simulated in Fig. \ref{NIIIandDIB}.}
         \label{HeII4686}
          \end{figure}

   \begin{figure}[t!]
  \centering
   \includegraphics[width=0.99\hsize]{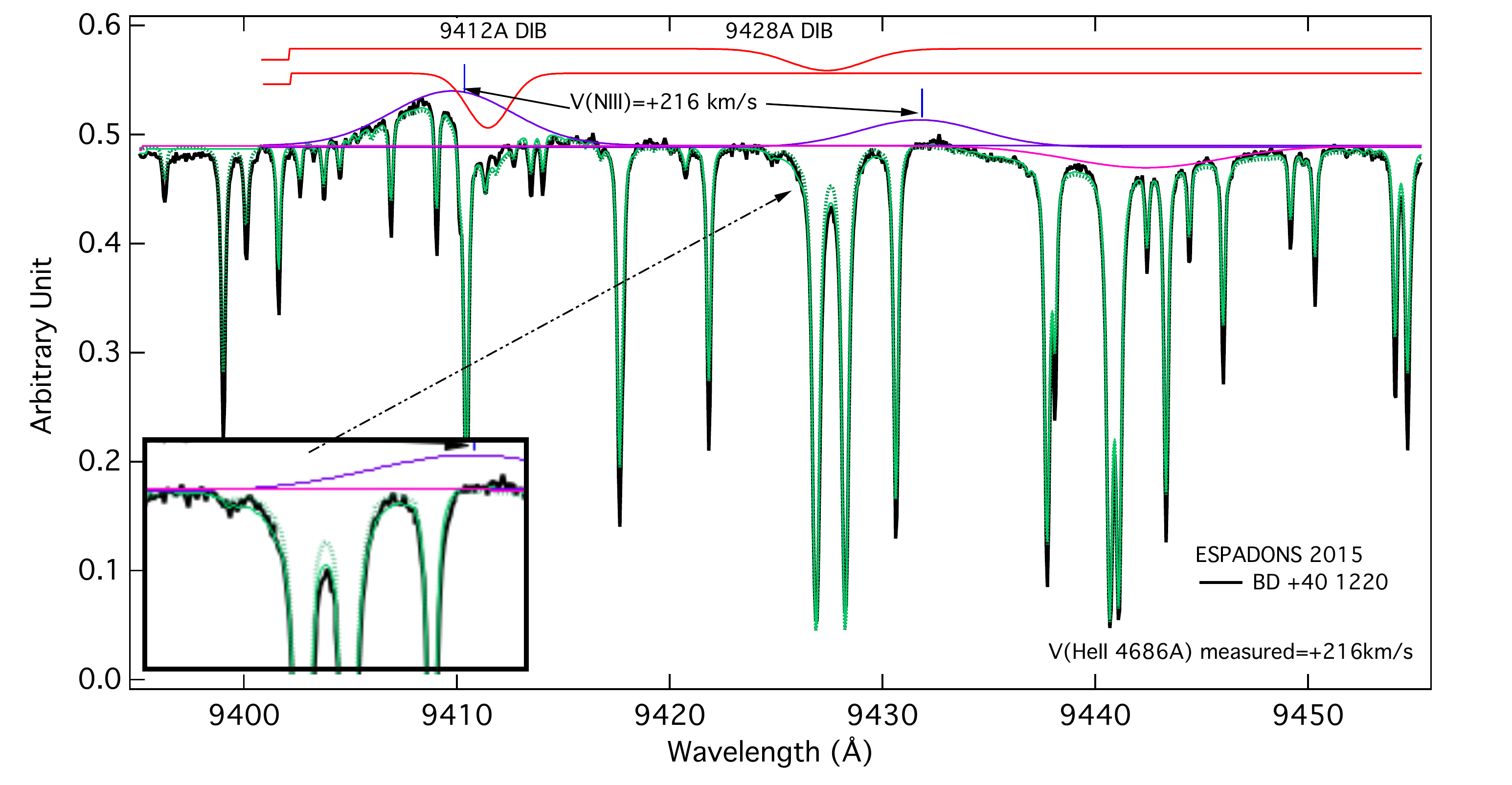}
      \caption{ESPaDOnS spectrum of BD+40 4220 (black line) and a simulation of the data (green line) based on a flat continuum, a convolved synthetic atmospheric water vapor profile, two Gaussian profiles representing the 9402.5 and  9424.5~\AA\ \ion{N}{iii} emission lines (violet lines), two Gaussian absorption profiles representing the two 9412 and 9428 ~\AA\ DIBs present in the spectral region (red lines), a broad stellar absorption line at $\sim$ 9443 ~\AA\ (pink line). The stellar emission lines have been Doppler shifted in agreement with the results shown in Fig. \ref{HeII4686}. The green dashed line is the same adjustment without the 9428 ~\AA\ DIB.}
         \label{NIIIandDIB}
   \end{figure}

An example is the case of the ESPaDOnS spectrum of Cyg~OB-2~5 (\object{BD+40 1220}) used in the aforementioned studies. This binary system has strong emission features associated with the colliding winds of the two members, as studied in detail by \citep{Rauw99}. The authors have published series of spectra recorded at all phases of the system. Fig.~\ref{HeII4686} shows the average of the profiles of the 4634 and 4640~\AA\ \ion{N}{iii} and 4686~\AA\  \ion{He}{ii} emission lines observed by \cite{Rauw99} for orbital phases 0.178 and 0.193, superimposed onto the ESPaDOnS spectrum. The emission line shapes are found to be consistent with the observations for such phases. Additionally, in order to match the data a Doppler shift of $+$216~km~s$^{-1}$ has to be applied, also consistent with the same range of orbital phase \citep[see Fig.~4 of][]{Rauw99}. In the NIR, a close inspection of the spectrum shows clearly an emission feature near 9410~\AA\  (Fig.~\ref{NIIIandDIB}). In this spectral region two \ion{N}{iii} transitions are present whose rest wavelengths are 9402.5 and  9424.5~\AA\ .  When the $+$216~km~s$^{-1}$ Doppler shift is applied these N\,{\sc iii} lines are shifted to 9409.3 and 9431.3~\AA\ respectively. The first one is at the location of the observed emission feature, and the second one is very close to the 9428~\AA\ DIB region. We do not possess information on the relative strengths of the two lines, however if the second line is present, as it is likely, it can partly mask the 9428~\AA\ DIB. Fig.~\ref{NIIIandDIB} shows an example of adjustment to the data using a combination of a flat continuum, a TAPAS synthetic profile convolved to the ESPaDOnS resolution \citep{Bertaux14_tapas}, two Gaussian emissions at 9409.3 and 9431.3~\AA\ representing the \ion{N}{iii} emission lines, a broad stellar absorption at 9443~\AA\, and finally two 5~\% and 2~\% deep Gaussian absorptions representing the 9412 and 9428~\AA\ DIBs, respectively. The \ion{N}{iii} lines are broadened with respect to the 4634 and 4640~\AA\ to take into account the longer wavelength. The very weak stellar absorption line at 9443~\AA\ seen here is also detected in other early-type star spectra. The depth of the 9428~\AA\ DIB is consistent with the strength of the observed 9577\AA\ DIB and laboratory ratios. The figure also shows the same adjustment in the absence of the 9428~\AA\ DIB. Although we do not demonstrate the existence of the DIB, the existence of such an adjustment clearly shows that it can not be precluded without better constraints on the actual emission lines in the DIB area.

\begin{figure}[t!]
  \centering
   \includegraphics[width=0.99\hsize,height=9cm]{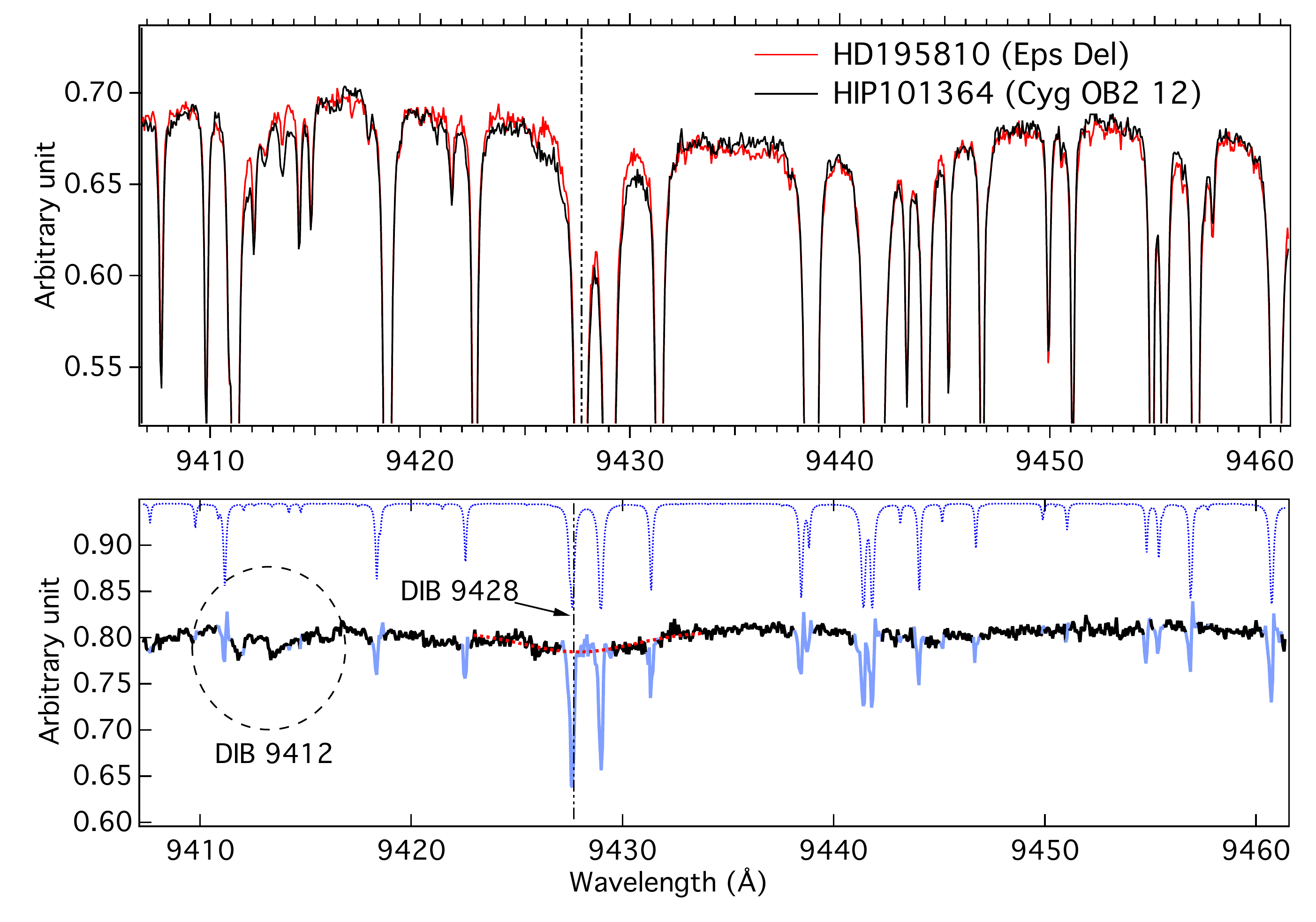}
      \caption{Top: Unnormalized CFHT ESPaDOnS spectra of nearby star HD\,195810 and the strongly reddened star HIP101364 in the 9428~\AA\ DIB region. No continuum fitting nor telluric correction has been performed and the spectra are displayed in the geocentric frame. Departures are seen at centers of telluric lines due to different atmospheric water vapor altitude profiles. Outside those lines departures can be seen around the 9412~\AA\ DIB \cite{Gala00,Cordiner17}. Very weak departures may also be seen in the 9428~\AA\ DIB region. Bottom: ratio between the two spectra. Strong telluric residuals are present (light blue regions), however two additional features are apparent: a complex structure around the 9412~\AA\ DIB and a depression around 9428~\AA. The dashed black vertical line indicates the expected position of the DIB based on laboratory results and the 9577~\AA\ DIB (not shown) in the same spectrum. A Gaussian fit performed in unmasked regions (dark blue). The depth of the 9428 ~\AA\ DIB is on the order of 2.5~\%, which should be consistent with the 7~\% depth of the 9577~\AA\ DIB.}
         \label{comp9428_espadons}
   \end{figure}

We have also revisited the search for the weak 9428\AA\ DIB for targets previously analyzed by \cite{Walker16} and \cite{Gala17}. To do so we have selected the ESPaDOnS spectra of the strongly reddened star \object{HIP101364} (\mbox{Cyg OB2 12}) and its comparison nearby star \object{HD\,195810} ($\epsilon$ Del) on the one hand, and recent EDIBLES spectra of both \object{HD\,169454} and the lower reddening \object{HD\,54662} on the other hand. The latter spectra have been recorded in very similar atmospheric conditions (PWV=1.25 and 1.34 mm respectively), and avoiding any adjustment for the airmass, at variance with the mentioned works. 
Our aim is to study the 9428~\AA\ spectral region without any continuum fitting nor telluric correction. We have simply compared the unnormalized spectra and their ratios in the geocentric frames. One disadvantage of this method is that DIBs in both targets are shifted due to the motion of the Earth and the different absorbing cloud radial velocities. However, if DIBs in the comparison star are weak, this should not be a problem. We show in Fig.~\ref{comp9428_espadons} (top) the unnormalized ESPaDOnS spectra of HIP\,101364 and HD\,195810 in the geocentric frame, as well as the ratio between the two spectra (bottom), after simple interpolation of the first spectrum over the same pixels as in the first one. The comparison between the two spectra reveals differences in the 9412~\AA\ DIB area. This is the location of the rather strong DIB first noticed by \cite{Gala00} and recently confirmed with STIS spectra by \cite{Cordiner17}. Other weak differences can be seen in the 9428~\AA\ DIB region. 

The ratio is characterized by strong amplitudes residuals at the telluric lines locations, and has two additional peculiar regions. First, there is a deviating feature with a complex shape in the 9412\AA\ DIB area, very likely due to this DIB. Of more importance here is the depression that can be clearly seen at the expected location of the C$_{60}^{+}$ 9428\AA\ DIB, despite the telluric residuals. A Gaussian fit using telluric masks shows that this depression is on the order of 2.5~\%, an order of magnitude in agreement with the 7~\% deep 9577\AA\ DIB for the same target. Although our method does not allow us to measure precisely the width and depth of this feature, but it shows that the absence of a 9428~\AA\ DIB in the HIP\,101364 spectrum is not proven. 

\begin{figure}[t!]
  \centering
   \includegraphics[width=0.99\hsize,height=6cm]{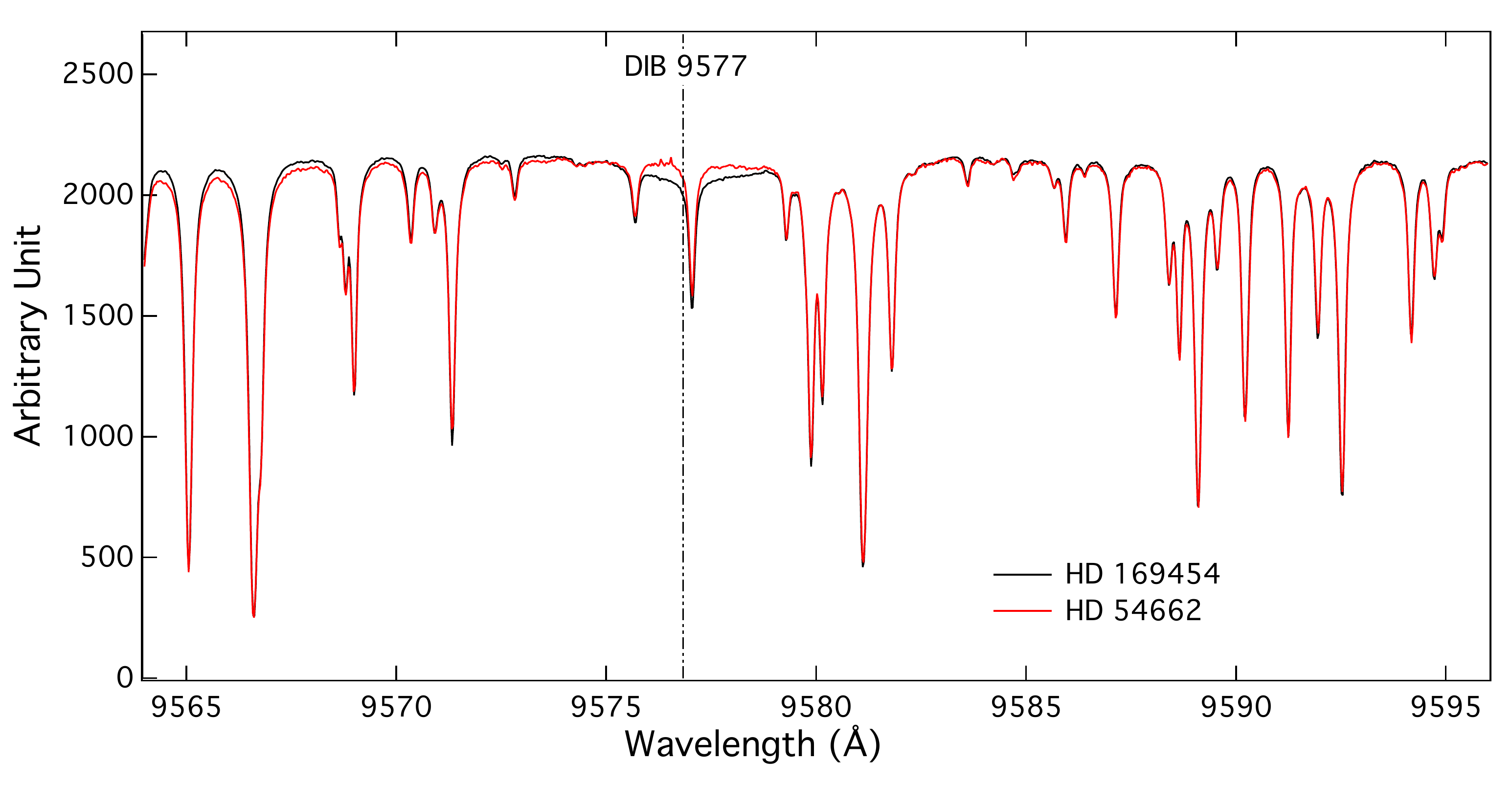}
      \caption{EDIBLES spectra of HD\,169454 and HD\,54662 in the 9577~\AA\: DIB region and in the geocentric frame. Apart from the difference between the strong DIB of HD\,169454 and the absence of a corresponding absorption in HD\,54662, the two spectra look similar, being characterized by similar atmospheric lines.}
         \label{stars169454_54662_9577}
   \end{figure}

\begin{figure}[t!]
  \centering
   \includegraphics[width=0.99\hsize]{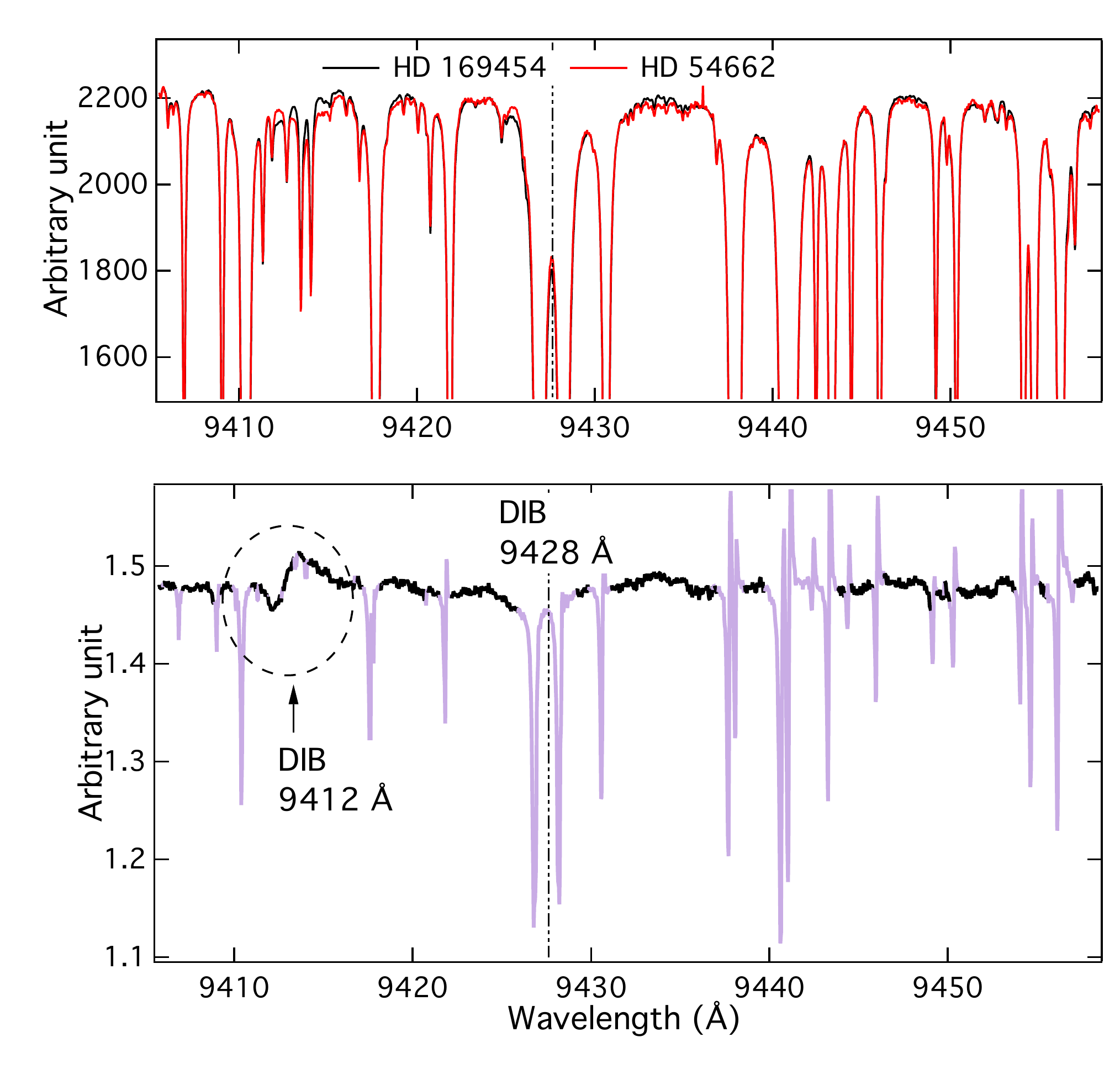}
      \caption{Top: Same as Fig.\ref{comp9428_espadons}, but for the two stars HD\,169454 and HD\,54662 (which were also shown in Fig. \ref{stars169454_54662_9577}). Departures are clearly seen around the 9412~\AA\: DIB reported by \cite{Gala00,Cordiner17}. Very weak differences are also seen in the 9428~\AA\: DIB region. Bottom: ratio between the two HD\,169454 and HD\,54662 spectra. Strong telluric residuals are present, however two additional features are apparent: a \textit{P Cygni} structure around the 9412~\AA\ DIB, resulting from the DIBs in both spectra, and a depression around 9428~\AA. The dashed black vertical line indicates the expected position of the DIB based on laboratory results and corresponds to the center of the depression.}
         \label{comp9428_edibles}
   \end{figure}

We have applied a similar simple method to HD\,169454. We first show in Fig.~\ref{stars169454_54662_9577} the EDIBLES spectra of the targets HD\,169454 and HD\,54662 in the 9577~\AA\: DIB spectral region. No continuum fitting has been performed, the two spectra are displayed with two different scales in such a way their upper parts match each other as closely as possible. The strength of the HD\,169454 9577~\AA\ DIB can be clearly seen, and the same DIB for HD\,54662 is too weak to be apparent on this graph. Fig.~\ref{comp9428_edibles} (top) shows the same spectra in the 9428~\AA\ DIB spectral region. All narrow features are telluric, as can be seen by comparison with a synthetic H$_{2}$O atmospheric profile displayed on top of the bottom graph, and  it can be checked that the telluric lines look all very similar\footnote{Synthetic telluric spectra have been downloaded from \texttt{http://cds-espri.ipsl.fr/tapas/} }. A close inspection of the figure reveals that at the expected location of the 9428~\AA\ DIB, the HD\,169454 spectrum is slightly below the one of HD\,54662, however it is difficult to conclude due to the weakness of the departures. On the other hand, there are clear departures around 9412~\AA\ . In this region the HD\,169454 spectrum is successively below and above the one of HD\,54662. The bottom graph shows the ratio between the two spectra, made after a simple interpolation of the HD\,54662 data over the same pixels as those of HD\,169454. There are strong residuals at the telluric line center locations, showing that the telluric lines do not match perfectly. Still, here they match enough to show that the ratio is about constant over the spectral interval outside telluric line central parts and in two locations, around 9412 and 9428~\AA\:  respectively. Around 9412~\AA\: there is  a \textit{P Cygni-type} structure. This is in agreement with the existence of the 9412\AA\ DIB in the two spectra. Because the spectra are displayed in the geocentric frame and the interstellar cloud radial velocities are different for the two stars, the DIBs are found spectrally displaced and the ratios produce the \textit{P Cygni-type} profile. The second feature located 9428~\AA\ is found to be centered close to the DIB wavelength (indicated in the figure) expected from (i) the C$_{60}^{+}$ central wavelength predicted by \cite{Spieler17} (i.e. 9427.5~\AA\ ), (ii) the HD\,169454 radial velocity of the main absorbing cloud (i.e. -10 km s$^{-1}$), (iii) the heliocentric correction during the HD\,169454 exposure (barycentric Earth radial velocity of $\sim$-14 km s$^{-1}$). The depth of the band measured on the graph is found to be on the order of 1.3~\% , a value fully compatible with the value measured by \cite{Walker16} and the depth of the 9577~\AA\ DIB shown in the Fig.~\ref{stars169454_54662_9577}, found to be on the order of 3.5~\%, and the EW and FWHM ratios measured recently by \cite{Campbell16b,Campbell16a} and \cite{Spieler17}, namely 1 vs 0.3 for the EW and 3.3 vs 2.5~\AA\ for the FWHM. Since HIP\,101364 and HD\,169454 are targets for which the existence of the 9428~\AA\ DIB has been precluded by \cite{Gala17}, we believe that the difficulty of the continuum fitting and telluric line correction is at the origin of this discrepancy. Since the present method does not use any correction and is made possible by very similar and good atmospheric conditions, it is more direct, and issues resulting form this difficulty are reduced.

\section{Conclusion and discussion}\label{sec:5}

We have considered with care the arguments used against the identification of the fullerene cation C$_{60}^{+}$ through its absorption bands. We have shown that the three main arguments may not be valid. First, there is an increased dependence on the stellar type of the 9632/9577 DIB ratio after its correction for stellar contamination \citep{Gala17}, showing that further analyses are needed to estimate properly the stellar contribution to the absorption in the 9632 band area. Second, we have shown that there is another interpretation for the double structure observed in the 9632~\AA\ region for the Orion star HD\,37022, according to which there is no discrepancy between the two strong DIBs 9577 and 9632~\AA\ for this star. Finally, we have shown that the previously claimed absence of 9428~\AA\ DIB in the spectra of BD+40 1220 (Cyg OB2 5), HIP\,101364 (Cyg OB2 12) and HD\,169454 is ambiguous. In the case of BD+40 1220, the existence of stellar emission lines due to colliding winds precludes any conclusion until these emissions are better characterized. For Cyg OB2 12 and HD\,169454 we have compared with spectra of low reddening targets characterized by similar telluric contamination and did not perform any adjustment for atmospheric water vapor columns. Doing so, we found evidence for the 9428~\AA\ DIB at the expected location and with the expected strength.

We have to combine these results with the recent findings and clarification about the 9365~\AA\ of \cite{GalaKre17}. The authors identified a new interstellar band around 9362~\AA\ and, taking into account this additional band, made several detections of absorptions at 9365~\AA\ , i.e. very close to the predicted wavelength \citep[e.g. 9364.9~\AA\ from][]{Spieler17}.  All together, this implies that the stronger four of the five bands may have been measured. Note that, according to all mentioned works, the fifth band is weak enough to have escaped detection most data analyses. 

This study shows that high signal spectra such like the EDIBLES data will significantly help in the study of the C$_{60}^{+}$ absorptions. It also shows that work on the modeling of weak NIR stellar lines of early-type stars is crucially needed. Such models will also become particularly useful in the perspective of future HST/STIS data acquired with the technique devised by \cite{Cordiner17}. In the absence of telluric contamination, the weak stellar contributions will be the ultimate but surmountable obstacle to the definitive identification of the five C$_{60}^{+}$ bands predicted from the laboratory experiments.

\begin{acknowledgements}
We thank our referee for his thorough review. M.E. acknowledges funding from the "Region Ile-de-France" through the DIM-ACAV project.
R.L. acknowledges support from "Agence Nationale de la Recherche" through the STILISM project (ANR-12-BS05-0016-02) and the CNRS PCMI national program. JC acknowledges support from an NSERC Discovery Grant.

Based on observations collected at the European Organisation for Astronomical Research in the Southern Hemisphere under ESO programme 194.C-0833 (EDIBLES Survey, PI. N.L.J.~Cox). 
Based on observations obtained at the Canada-France-Hawaii Telescope (CFHT) under programme 05AO5 (PI. B.H.~Foing).
CFHT is operated by the National Research Council of Canada, the "Centre National de la Recherche Scientifique" of France, and the University of Hawaii.
This research has made use of the SIMBAD database, operated at CDS, Strasbourg, France.

\end{acknowledgements}
\bibliographystyle{aa} 
\bibliography{mybib.bib}
\end{document}